\documentclass{article}
\usepackage{graphicx}
\usepackage{comment}
\usepackage[margin=1.5in]{geometry}
\usepackage{url}            % simple URL typesetting
\usepackage{booktabs}       % professional-quality tables
\usepackage{amsfonts}       % blackboard math symbols
\usepackage{nicefrac}       % compact symbols for 1/2, etc.
\usepackage{microtype}      % microtypography
\usepackage{indentfirst}
\usepackage{color,ulem}

\usepackage{authblk}

\title{Automatic Ontology Learning from Domain-Specific Short Unstructured Text Data}
\author[1]{Yiming Xu}
\author[2]{Dnyanesh Rajpathak}
\author[3]{Ian Gibbs}
\author[4]{Diego Klabjan}
\affil[1]{Department of Statistics, Northwestern University}
\affil[2]{Research and Development, General Motors}
\affil[3]{Vehicle Safety, General Motors}
\affil[4]{Department of Industrial Engineering and Management Sciences, Northwestern University}

\begin{document}
\maketitle

\begin{abstract}
Ontology learning is a critical task in industry, dealing with identifying and extracting concepts captured in text data such that these concepts can be used in different tasks, e.g. information retrieval. Ontology learning is non-trivial due to several reasons with limited amount of prior research work that automatically learns a domain specific ontology from data. In our work, we propose a two-stage classification system to automatically learn an ontology from unstructured text data. We first collect candidate concepts, which are classified into concepts and irrelevant collocates by our first classifier. The concepts from the first classifier are further classified by the second classifier into different concept types. The proposed system is deployed as a prototype at a company and its performance is validated by using complaint and repair verbatim data collected in automotive industry from different data sources. 
%\keywords{Ontology learning \and Classification \and Information retrieval}

\end{abstract}

\section{Introduction} 
Over 90\% of organizational memory is captured in the form of unstructured as well as structured text data. The unstructured text data takes different forms in different industries, e.g. body of email messages, medical records of patients, contracts, fault diagnosis reports, speech-to-text snippets, call center data (e.g. conversation, notes, replies, etc.), design and manufacturing data, data generated over different social medical platforms, etc. Given the ubiquitous nature of unstructured text data collected in industries, they provide a rich source of information. For example, in automotive which is used as our running example or aerospace industry in the event of fault or failure, the repair verbatims (commonly referred to as verbatim) are captured during fault and root cause investigation \cite{Rajpathak12}. These repair verbatims capture valuable information indicating the nature of fault, the possible root causes behind such faults, and the corrective actions taken to repair the faults. Such type of knowledge when extracted provides a critical insight into ways the parts or components fail during their usage and under different operating conditions and the root causes associated with the faults. This knowledge provides useful information to business to improve the quality of product to ensure avoidance of similar faults in the future. However, efficient and timely extraction, acquisition, and formalization of knowledge from unstructured text data poses several challenges: 1. the overwhelming volume of unstructured text data makes it difficult to manually extract the critical concepts embedded in the data, 2. the lean use of language and vocabulary results into inconsistent vocabulary, e.g. ¡®vehicle¡¯ vs ¡®car,¡¯ or ¡®failing to work¡¯ vs. ¡®inoperative,¡¯ etc., and finally, 3. different types of noises are observed in unstructured text data, e.g. misspellings, run-on words, additional white spaces, and abbreviations.

An ontology \cite{Gruber} provides an explicit specification of concepts and resources associated with domain under consideration. A typical ontology consists of concepts and their attributes commonly observed in a domain, relations between such concepts, a hierarchy representing how such concepts are related to each other, and concept instances representing ground-level objects. For example, the concept ¡®vehicle¡¯ can be used to formalize a locomotive object and vehicle instances, such as ¡®Chevrolet Equinox.¡¯ An ontological framework and the concept instances can be used to share the knowledge among different agents in a machine-readable format (e.g. RDF/S \footnote{https://www.w3.org/TR/rdf-schema/}) and in an unambiguous fashion. Hence, ontologies constitute a powerful way to formalize the domain knowledge for supporting different application, e.g. natural language processing \cite{Cimiano} \cite{Girardi}, information retrieval \cite{Middleton}, information filtering \cite{Shoval} among others. 

In this work, we propose an approach to automatically learn a domain-specific ontology. The process of ontology learning is divided into two phases: 1. a classifier is trained to classify the collocates in a verbatim into concepts and irrelevant collocates and 2. concepts are further classified into specific types. Note that a concept is also a collocate. We build a two-stage classification framework rather than single stage because concepts have the notion of types while irrelevant collocates have no such notion. Our input consists of a corpus made of short text or verbatim and an incomplete set of concepts and their types in the domain under consideration. The goal is to identify further concepts and their types. Note that a collocate can be a concept in one verbatim but an irrelevant collocate in another verbatim. For the first step, we use the concepts in the incomplete seed ontology as positive samples, while we develop a new algorithm to create negative samples. In our classification models, we use both linguistic features (POS, etc.) and word embedding features (word2vec). Polysemy pose a significant problem since they occur frequently in short text. We capture them as features in our models as follows. Given a 1-gram, we cluster the embedding vectors of collocates with the number of clusters equal to the number of polysemy of the 1-gram based on WordNet \cite{Miller}. Given an occurrence of the 1-gram, we use as a feature the centroid of the closest cluster. We have also done two rounds of active learning to augment the training set. 

The key contributions of our work are as follows. 1. In several domains, the collocates of different sizes are not homogeneous and initially we develop a single classifier to classify the collocates of different sizes, which provided only limited accuracy. To overcome this problem, we identify a common set of features associated with collocates of different sizes and utilize them to develop different classifiers where each classifier corresponds to each category of the collocate. 2. The problem of polysemy is ubiquitous in text corpora. We resolve the problem of polysemy by clustering word embeddings and assigning a 1-gram into an appropriate cluster that is used as a feature. 3. In real-world data, it is common to find abbreviations. We resolve the problem of abbreviation disambiguation and to the best of our knowledge we are the first domain-specific proposal to disambiguate abbreviations by combining a statistical and machine learning approach. 4. From the engineering perspective, ours is a practical ontology learning system, which is successfully employed as a proof-of-concept tool for extracting ontologies from real-world data at an industrial scale.

The rest of the paper is organized as follows. In the next section, we provide a review of the relevant literature. In Section 3, we provide the problem description and an overview of our approach. In Section 4, we discuss our model specifications, data preprocessing algorithms and features used in classification. In Section 5, we discuss in detail the experiments and evaluation of our classification models. And finally, in Section 6, we conclude our paper by reiterating the main contributions.

\section{Background and Related Works}

A plethora of works have been done in ontology learning. There are three major approaches: statistical methods (e.g. weirdness, TF-IDF, etc.), machine learning methods (e.g. bagging, Na?ve Bayes, HMM, SVM, etc.), and linguistic approaches (e.g. POS patterns, parsing, WordNet, discourage analysis, etc.). 

\cite{Wohlgenannt} built an ontology learning system by collecting the evidences from heterogeneous sources in a statistical approach. The candidate concepts were extracted and the ¡®is-a¡¯ type of relations were constructed by using chi-square co-occurrence significance score. In comparison with \cite{Wohlgenannt}, we use a structured machine learning approach so that it can be applied on unseen data. In \cite{Wohlgenannt}, all evidences were integrated into a big semantic network. The spreading activation method was then employed to find the most important candidate concepts. The candidate concepts were then manually evaluated before adding to an ontology. In comparison with this, in our approach, the machine learning based methods are employed to make use of the latent features embedded in the text data to automatically learn the ontology and very limited human interventions are needed in our approach. The only human involvements in our work are active learning and manually labeled training collocates, which are not necessary but just a way to augment training data. Moreover, since our approach makes use of different features identified from the data, e.g. context features, polysemy features, etc., it exploits richer data characteristics compared to \cite{Wohlgenannt}. Finally, ours is a probabilistic model that considers the context of a concept, which allows us to handle the extraction of unseen phrases effectively. The model proposed by \cite{Wohlgenannt} is deterministic in nature, which does not consider the context and therefore fails to handle previously unseen phrases.
 
\cite{Doing-Harris} made use of cosine similarity, TF-IDF, a so-called C-value statistic, and POS to extract the candidate collocates for constructing an ontology. This work is done in a statistical and linguistic approach. \cite{Yosef} constructed a hierarchical ontology by employing support vector machine that heavily relies on using POS as the primary feature to determine the classification hyperplane boundary in a machine learning and statistical approach. While POS is an important feature in text data, as we have discussed, additional features such as the word embedding features further boost the performance of the classifier. Such features consider the context in which specific phrases are referenced. As word embedding features were not considered by \cite{Yosef} it is difficult to envisage how the context associated with each concept was considered while extracting the relevant phrases. \cite{Pembeci} evaluated the effectiveness of word2vec features in ontology construction, which used the statistic based on 1-gram and 2-gram counts to extract the candidate concepts. The ontology was then constructed manually. In our work, we not only consider word2vec as one of the features but other critical features such POS, polysemy features, etc., are utilized to provide us with a robust probabilistic machine learning model. We do not use statistical features from other works since word embedding features dominate statistical features.

\cite{Ahmad} constructed an ontology by using the ¡®weirdness¡¯ statistic to extract candidate concepts. Next, the collocation analysis was performed coupled with domain expert verification to extract the ontology. There are two key differences between our approach and the one proposed by \cite{Ahmad}. First, our concept extraction classifier uses the existing ontology, stop words list, and noise words list as the support mechanism to train the model, while in their approach the notions of ¡®weirdness¡¯ and ¡®peakedness¡¯ statistics were used to extract the candidate concepts. And second, in their work, there was a heavy reliance on domain experts to verify the newly constructed ontology, whereas no such manual involvement is needed in our framework either during concept extraction or classification stages. As very few of human involvements are needed in our approach, it can be deployed as a self-sustained algorithm to learn an ontology in a specific domain. 

In our work, we also propose a new approach to disambiguate abbreviations. There are several related works. \cite{Stevenson} extracted features such as concept unique identifiers and then built a classification model. \cite{HaCohen} identified context based features for classification, but they assumed an ambiguous phrase only has one correct expansion in the same article. \cite{Li} proposed a word embedding based approach to select the expansion from all possible expansions with largest embedding similarity. There are two major differences between our approach and these works. First, we propose a new model which combines the statistical approach (TF-IDF) and machine learning approach (Naive Bayes model) together, i.e. we measure the importance of each collocate by TF-IDF and estimate the posterior probability of each possible expansion, while in their work they either only applied machine learning classification models or simply calculated statistical similarity between abbreviation and possible expansions. Second, in these works strong assumptions were made, such as each phrase only has one expansion in the same article and features are conditionally independent, while we do not have any assumptions for our model and therefore it is more robust.

\section{Problem Statement and Approach}

In industry, the data from several sources provides valuable information. However, given the overwhelming size of real-world data, it is usually impossible to manually go through the data to discover all the critical information. In this work, we focus on unstructured short text data. Here is a verbatim from the automotive industry: `Customer states the engine control light is on with 10. The dealer identified internal short to the fuel pump relay. The engine control module is replaced and reprogrammed and the DTCs are cleared.' Our aim is to extract concepts from the data, such as ¡®engine control module,¡¯ ¡®fuel pump relay,¡¯ ¡®internal short,¡¯ ¡®replaced and reprogrammed,¡¯ etc., from each verbatim and further classify these concepts into different types. For example, ¡®engine control module' is a part concept and ¡®replaced and reprogrammed¡¯ is an action concept. 

In addition to the corpus which contains millions of verbatims, we are also given an incomplete set of concepts and their types as part of an ontology. The concepts do not have the corresponding verbatim and thus we assume that every occurrence of a concept from the ontology is indeed a concept (regardless of the underlying verbatim). 

%Formally, we define our two-stage classification problem as follows:

%Let $C$ be a finite set of ontological concepts (also referred to as the classes), $W$ be a set of n-gram collocates, and $T$ be a finite set of taxonomical relations between the n-gram collocates. In our domain ontology, $C$ can be denoted as \{$D, F, Concept, Irrelevant Term, Concept Type 1,$ 
%$Concept Type 2, ...$\}, where $D$ represents a specific domain, $F$ represents the features used to classify $W$, $Concept$ and $Irrelevant Term$ represent the initial classification of $W$ into \{Concept, Irrelevant Term\} classes respectively, and finally $Concept Type 1, Concept Type 2, ...$ represent the classes assignments to the $Concept$ n-gram collocates. Our classifiers are then $f: W \to \{Concept, Irrelevant Term\}$, which assigns W to {Concept, Irrelevant Term}, and $f': Concept \to \{Concept Type 1, Concept Type 2, ...\}$, which assigns $Concept$ to $\{Concept Type 1,$ 
%$Concept Type 2, ...\}$. 

Since a text corpus usually contains different types of noises, we start by cleaning the verbatims; misspelling correction, run-on words correction, removal of additional white spaces, and abbreviation disambiguation are employed to clean the data. Then we tag the collocates of the verbatim phrases as concepts and irrelevant collocates to create the training set. We use the incomplete ontology and a few additional manually identified concepts. The irrelevant collocates are designed as the collocates between concepts and with certain additional properties detailed in Section 4. Next, the non-descriptive stop words in our data are deleted as they do not add any value to the task of classification.

Having deleted the stop words, the collocates are collected to provide necessary coverage to the different lengths of the concept phrases, e.g. ¡®Battery,¡¯ ¡®Fuel Cell,¡¯ ¡®Engine Control Module,¡¯ ¡®Inoperative,¡¯ ¡®Rough Surface to Finish,¡¯ etc. As we discuss in Section 4, we identify several unique features related to concepts and irrelevant collocates, such as POS, polysemy features, etc. 

Next, by using the labeled data and the features identified from the data a classification model is trained to classify collocates extracted from the verbatim into concepts and irrelevant collocates. The concept collocates and the features are then fed to the second stage classification model which assigns types to them. In the concept and irrelevant collocate classification, we have conducted two rounds of active learning to improve the performance. In inference, the model takes raw verbatims and preprocesses them using our pipeline, and then it extracts all candidate concepts without stop words and noise words. Finally these concept candidates are fed as input to our two-stage classification system. Figure 1 shows the overall process of our two-stage classification system.
\begin{figure}[h]
    \includegraphics[width=0.48\textwidth]{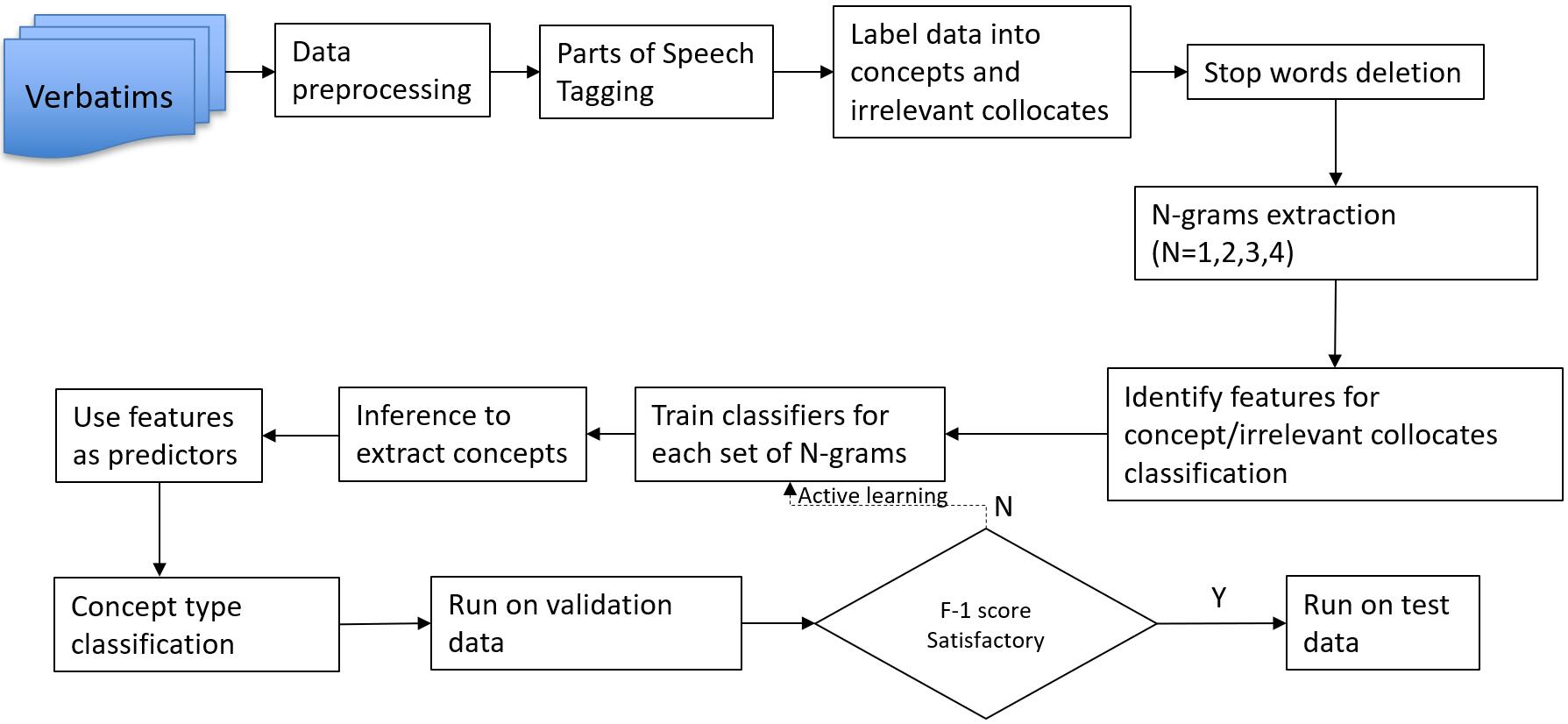}
    \centering
  \caption{The overall flow of the two-stage classification model.}
  \centering
  \end{figure}

\section{Model Specifications}

As discussed in the previous section, we construct the ontology by means of extracting concepts from the unstructured text verbatims and determine whether the candidates are concepts or irrelevant collocates. Several steps require word2vec which is trained on verbatims. We then further assign classes to the concept collocates. However, before extracting candidate concepts, data preprocessing is needed due to several types of noises.

%\subsection{Two-stage classification system}

\subsection{Data preprocessing}
To handle the noises, we define the correctness of a collocate by membership in the English dictionary or the existing ontology. The following steps are undertaken.

\textbf{1. Misspellings correction.}
We consider all possible corrections of a misspelled 1-gram each with Levenshtein distance of 1. If there is only one correction, we replace the misspelled 1-gram by the correction. Otherwise, for each candidate correction we define its similarity score to be the product of its logarithm of frequency and the word2vec similarity between the misspelled 1-gram and its correction. The misspelled 1-gram is replaced by the correction with the maximum similarity score.

\textbf{2. Run-on words correction.}
We split run-on words into a 2-gram by inserting a white space between each pair of neighboring characters. For a specific split, if both the left 1-gram and the right 1-gram are correct, we split the run-on 1-gram in such way. If there are multiple possible splits with correct 1-grams, then for each correct split, we define its similarity score to be the maximum of word2vec similarities between the run-on 1-gram and the two 1-grams. The split with maximum similarity score is replaced as the correct split.

\textbf{3. Removal of additional white spaces.}
We also observe several cases in the data where there are additional white spaces inserted in a 1-gram, e.g. ¡®actu ator.¡¯ We try to remove the additional white spaces to see whether it turns the two incorrect 1-grams into a correct 1-gram and if it does, then we employ this correction.

\textbf{4. Abbreviation disambiguation.}
Finally, it is ubiquitous in a corpus that same abbreviations have different expansions. It is critical to disambiguate the meaning of abbreviated collocates. Typically, an abbreviation is a collocate that can be mapped to more than one possible expansion (or full form), for example, `TPS' could stand for `Tank Pressure Sensor,' `Tire Pressure Sensor' or `Throttle Position Sensor.' The abbreviations mentioned in our data are identified by using the domain specific dictionary, which consists of commonly observed abbreviations and their possible full forms. For an identified abbreviation with a single full form, we replace that specific abbreviation with its full form. Otherwise we employ the following model.

Suppose an abbreviation $abbr$ has $N$ possible full forms, namely, $\{ff_1, ff_2, ..., ff_N\}$, where $N > 1$. We first collect the 1-gram collocates, which are co-occurring with $abbr$ from the entire corpus. The context collocates co-occurring with $abbr$ are denoted as $C_{abbr}$ and the set of all co-occurring collocates for $ff_n$, $1 \leq n \leq N$ are denoted as $C_n$. To prevent meaningless expansions and to compare the posterior probabilities of $ff_i$ and $ff_j$, we only focus on the intersection of these sets: $V =  \cap^N_{n=1} C_n \cap C_{abbr}$. Having identified the relevant intersecting collocates, we measure the importance of each collocate in terms of TF-IDF. 

Let $v_u\in \mathbb{R}^{|V|}$ be the TF-IDF vector of collocate $u=abbr$ or full form $u=ff_i$. Given that $ff_n$ is associated with $abbr$, the probability of co-occurring concepts given $ff_n$ is then estimated as $P(abbr|ff_n)=\prod_{i=1}^{|V|}(\frac{v_{ff_n,i}}{\sum \limits_{j=1}^{|V|}v_{ff_n,j}})^{v_{abbr,i}}.$ The intuition for this formula is that if $abbr$ and $ff_n$ are interchangeable, then they have the same distribution of co-occurring concepts. Therefore, we estimate the probabilities of the co-occurring concepts of $abbr$ with those from $ff_n$. Furthermore, we estimate the prior probability $P(ff_n)$ of $ff_n$ from its document frequency. Therefore, by the Bayes theorem, $P(ff_n|abbr)\propto P(abbr|ff_n)\cdot P(ff_n)$. We then replace abbreviation $abbr$ by the full form with the largest posterior probability.

\subsection{Preparation of training set}

The process of classifying the data into concepts and irrelevant collocates starts by labeling some of the raw collocates in order to create the training set. Given the scale of real-world data, it is impossible to manually label each raw sample collected. To overcome this problem, we rely on the existing incomplete ontology. We tag the data by using the incomplete ontology which tags all occurrences of concepts in the incomplete ontology as concepts. The collocates that are not tagged by the incomplete ontology are potentially irrelevant collocates or the concepts not covered by the incomplete ontology. For the purpose of avoiding repetitions and keeping concepts as complete as possible, only the longest collocate is marked as a concept. For example, if ¡®engine control module¡¯ is marked as concept, then its subgrams such as ¡®engine¡¯ or ¡®module¡¯ are not labeled as concepts. The remaining collocates that are not tagged by the incomplete ontology are extracted and labeled as irrelevant collocates.

However, since the size of the incomplete ontology is limited, it is of great importance to augment the current training set. We collect all of the collocate in verbatims which meet some frequency criteria. These frequently appeared collocates are then manually labeled to augment the training data.

In inference, given a verbatim, we collect all possible collocates without stop words and noise words in them, then these collocate candidates are passed to the two-stage classification system.

%\subsubsection*{Concepts labeling}
%In order to build classifiers to classify the extracted candidates into concepts/irrelevant terms, we need to first obtain the labeled data. To obtain labeled concepts and irrelevant terms, we first collect all the concepts with term frequencies greater than a lower bound, for example 100, to avoid outliers. Then we use the existing taxonomy as a criterion: if a candidate exists in the taxonomy, we label it as concept. For the purposes of avoiding repetitions and keeping concepts as complete as possible, our rule is that if longer candidate is labeled as concept, then its sub-concepts are not considered as concepts. 

%For example, if `engine control module' is labeled as concept, we do not consider `engine control' as concept. 

%\subsection{Concept/irrelevant term classification}

%\subsubsection*{Concept/irrelevant term classifiers}
%Using the extracted candidate concepts with their features, we train the following classifiers: Naive Bayes, XGBoost and random forest. Since it is more reasonable to treat concepts of different lengths separately, in our work, there is one classifier associated with each N-gram, where $N=1,2,3,4$, so that the model does not confuse concepts of different lengths.

%the process: we first tried ... then ... we realized that... 
%... but randomforest works the best...
%

%\subsection{Concept types classification}
%We further classify the concepts predicted from previous steps into different concept types. The same features as concept/irrelevant term extraction are applied in this step. 

\subsection{Feature engineering}
In our model, different features such as discrete linguistic features, word2vec features, polysemy centroid features, and finally the context based features are identified. To this end, we are given a collocate (either labeled as concept or irrelevant) and the underlying verbatim. These features are discussed in detail next.

\textbf{1. Discrete linguistic features.}
The following linguistic features are recognized: 1) POS related to each collocate identified by employing Stanford parts of speech tagger \cite{Ratnaparkhi}, 2) the POS tags of the three nearest left side 1-grams of the collocate, 3) the POS tags of the three nearest right side 1-grams of the collocate, 4) the POS tag of the nearest concept on the left side of the collocate, 5) the POS tag of the nearest concept on the right side of the collocate.

\textbf{2. Word2vec features. }
We also consider the continuous word2vec vector associated with each collocate as one of the features to improve the performance of the model. We train a Skip-Gram model with respect to frequent 1-grams. When the word2vec embedding is not available, we consider it as a zero vector. For a collocate, the associated feature vector is the average word2vec embedding of all of its 1-grams.

\textbf{3. Context features.}
We consider the `context' word2vec feature of each collocate. For a collocate $T$, we take the 3 left 1-grams and 3 right 1-grams of $T$ in its verbatim and obtain the word2vec embeddings of these 6 1-grams. The context feature is the concatenation of the average of the 3 embedding on the left and the average on the right. If a collocate is toward the beginning or the end of the verbatim and thus has less than 3 embeddings, then all missing are not considered in the average. If none is present, we set the average to be the zero vector. 

\textbf{4. Polysemy centroid features.}
Furthermore, we also consider the polysemy of a 1-grams. We employ the following two steps as shown in Figure 2. 1. For each collocate $T$, we take 1,000 verbatims in which $T$ is mentioned and calculate the context feature vector $V(T)$ for $T$ in each selected verbatim. Then we use WordNet to obtain the number $p$ of polysemies of $T$. Further, we use the k-Means algorithm to cluster these 1,000 $V(T)$ vectors, with the number of clusters set to $p$. 2. Having these polysemy centroids ready to use, for a collocate $T'$, we find the context vector from its verbatim. The feature vector of $T'$ corresponds to the closest centroid among those obtained in step 1 for $T'$ with respect to the context features of $T'$.
 
\begin{figure}[h]
    \includegraphics[width=0.41\textwidth]{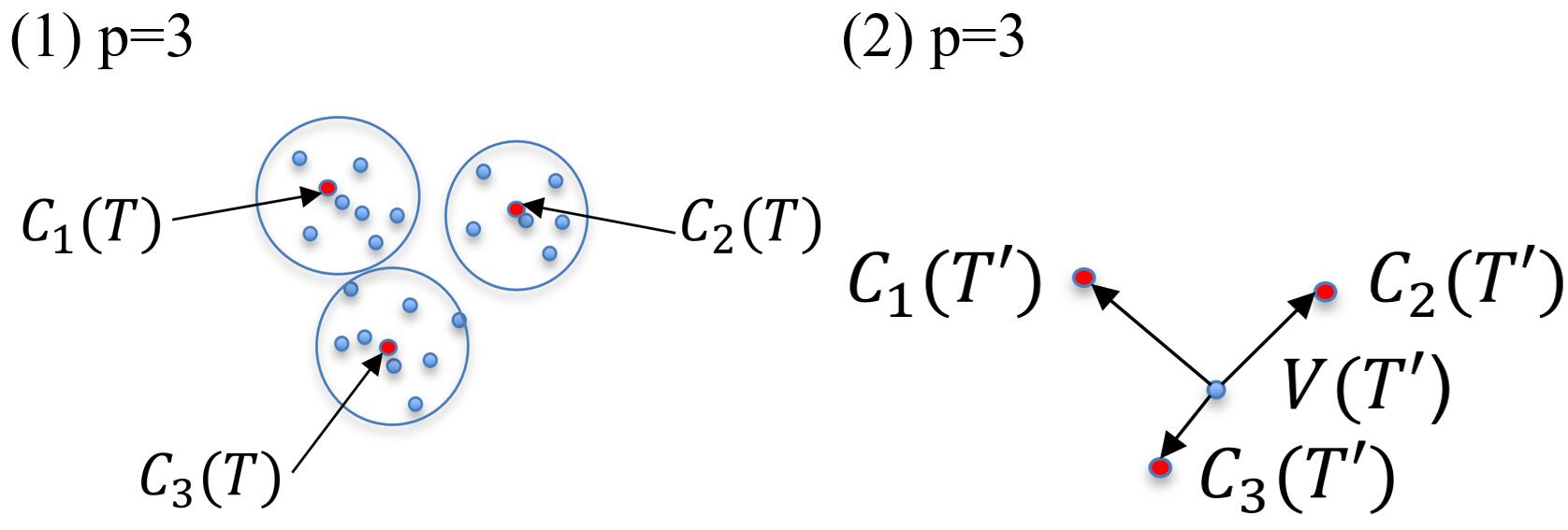}
    \centering
  \caption{(1) Obtain all possible polysemy centroids of a collocate: for a collocate $T$, we cluster context vectors and save the cluster centroids $C_1(T),...,C_p(T)$. (2) Create polysemy centroid feature of a collocate: for a new collocate $T'$, let $m=argmin\{d(V(T'),C_1(T')),...,d(V(T'),C_{p'}(T'))\}$ denote the index of the closest centroid, where $d$ is the Euclidean distance. Vector $C_m(T^\prime)$ is our polysemy feature for $T'$.}
 \centering
  \end{figure}

\textbf{5. Features based on the incomplete ontology.}
We also find that the incomplete ontology plays a significant role in classification. For a collocate, we split it into 1-grams, and add a feature vector of the same length as the collocate, with each element being set to be 1 if this 1-gram exists in the incomplete ontology, otherwise 0.

\subsection{Classification}
We use the random forest model as the classification model. We have also experimented with support vector machine, XGBoosting etc., but the experiments showed that random forest outperforms other models. We have fine-tuned the following important hyperparameters in random forest: the number of trees in the forest is 10, no maximum depth of a tree, the minimum number of samples required to split an internal node is 2. 

To improve the model performance, we have also introduced active learning to augment the training data. We train 8 different classifiers and feed randomly sampled unlabeled data to these classifiers. We gather the samples with 4 positive and 4 negative votes from the eight classifiers. We manually label all such samples, which the classifiers fail to classify into their correct classes due to the disagreements among them. All manually labeled samples are added back into the training set and the process is repeated twice. 
%The process of active learning used in our approach is shown in Figure 2.

%\begin{figure}[h]
%    \includegraphics[width=0.5\textwidth]{4.jpg}
%    \centering
%  \caption{Active learning process through training of classification %model.}
%  \centering
%  \end{figure}
  
We also analyze feature importance. We use a backward elimination process in which we initially start with all features and then drop one feature at a time and train our model by using the remaining features. This is done for all features. Then we remove the feature that yields the largest improvement to the F1-score when removed. This process is repeated iteratively until removing any feature does not improve the F1-score. The final set of features kept are Word2vec, Polysemy, POS, Context and Existing Ontology, which are the most important features in our model. The features dropped are left POS, right POS, left three POSes, right three POSes.

\section{Computational Study}
The ontology learning system is validated on a subset of an automotive repair (AR) verbatim corpus collected from an automotive original equipment manufacturer, the vehicle ownership questionnaire (VOQ) complaint verbatim collected from National Highway Traffic Safety Administration\footnote{https://www.nhtsa.gov/} and Survey data. The AR data contains more than 15 million verbatims, each of which on average contains 19 1-grams. Here is a typical AR verbatim: `c/s service airbag light on. pulled codes 100 \& 200..solder 8 terminals on both front seats as per special policy 300b.clear codes test ok.' The classification models are trained on AR. To study the generality of our model, we also test on VOQ which contains more than 300,000 verbatims where the verbatims are significantly different from AR. In VOQ, the issues are reported directly by customers, and thus it is more verbose in nature while AR is more technical and professional. A sample from VOQ reads: `heard a pop. all of the sudden the car started rolling forward....' Finally, the Survey dataset is generated by selecting the repair verbatims associated with different failures from AR and thus they look similar. Our incomplete ontology contains slightly more than 9,000 collocates and has three types labeled as A,B,C herein.

The classification system is implemented in Python 2.7 and Apache Spark 1.6 and ran on a 32-core Hadoop cluster. We evaluate the performance of abbreviation disambiguation and the performance of the first stage and second stage classification models.

\subsection{Evaluation of abbreviation disambiguation}

%For the first experiment we randomly select raw verbatim from three data sources, i.e. GART, the vehicle ownership questionnaire(VOQ) complaint verbatim collected from National Highway Traffic Safety Administration\footnote{https://www.nhtsa.gov/}, and survey data. The data are fed to the algorithms and initially the algorithm identifies incorrect words by our definition. All the incorrect terms are given as the input first to run-on words correction algorithm followed by removal of additional white space algorithm, and finally to the misspellings correction algorithm. The order of these algorithms is determined empirically. Table 1 summarizes the findings of the algorithms in correcting the incorrect terms.

To evaluate the performance of the abbreviation disambiguation algorithm, we generate three test datasets from the AR data source. On average, 5\% of AR verbatims contain an abbreviation, and each abbreviation has more than 2 expansions. Table 1 summarizes the results of the abbreviation disambiguation algorithm experiment, which are manually evaluated by domain experts. 

\begin{table}[h]
\centering
      \caption{The result summary of abbreviation disambiguation algorithm. $N_{raw}$ denotes the number of raw verbatims, $N_c$ denotes the number of abbreviations corrected and $N_{correct}$ denotes the number of correct abbreviation corrections.}
  \begin{tabular}{|l | l |l|l|l | c |}
  \hline
     \textbf{Data}  & \textbf{$N_{raw}$} & \textbf{$N_c$} & \textbf{$N_{correct}$}&\textbf{Accuracy} \\ \hline
AR 1 &10,000& 204&154&0.75\\\hline
AR 2 &30,000& 374&278&0.74 \\\hline
AR 3 &45,000& 407&312&0.77 \\\hline
  \end{tabular}
\end{table}
As it can be seen in Table 1, the performance of the algorithm is stable, i.e. the accuracies do not vary much on the three test datasets. On average, 75\% of our corrections are correct, which shows our algorithm is able to capture correct expansions of abbreviations. Note that there might be abbreviations that are not captured by our algorithm if abbreviations are not in the abbreviation list.

\subsection{Performance of classifiers}
Recall that the training data is from the AR data. Since the entire AR data is large, our training set is sampled from AR in the following way: for each N-gram ($N=1,2,3,4$), we randomly take 50,000 concepts and 50,000 irrelevant collocates which we regard as the training set for the N-gram model. Among them, 2,000 are manually labeled and 2,000 are from active learning. For evaluation, we generate three different test datasets. The first test dataset consists of 3,000 randomly selected repair verbatim from the AR data but only 1,500 extracted candidate collocates are manually evaluated. The second test dataset consists of 23,000 verbatims from VOQ, and 1,500 candidate collocates are manually evaluated. The third test dataset (Survey) is generated by selecting the repair verbatim associated with different failures from AR. It consists of 46,000 verbatims and 1,000 extracted candidate collocates are manually evaluated. These data are preprocessed by the data preprocessing pipeline and the cleaned data are used in inference. In the AR test set, from 3,000 verbatims, the algorithm extracts 39,000 concepts and irrelevant collocates. Among those classified as concepts, slightly less than 30\% are previously unseen unique concepts, which verifies that our system is efficient in learning an ontology. After obtaining all extracted concepts and their types, we then randomly select a subset of the results which are given to three Subject Matter Experts (SMEs). The SMEs read the actual verbatims to identify the context and then they manually evaluated each classification result. The Precision, Recall, and F1-score for the test datasets based on the labels marked by the SMEs are given in Table 2. 

\begin{table}[h]\centering     
\caption{The evaluation of concepts and irrelevant collocates classification algorithm.}
  \begin{tabular}{|l | c |c|c|}
  \hline
     \textbf{Dataset}  &\textbf{Precision}&\textbf{Recall}&\textbf{F1-score}  \\ \hline
AR &0.81&0.90& 0.85\\\hline
VOQ & 0.89 &0.47&0.62 \\\hline
Survey & 0.80 &0.79&0.79 \\\hline
  \end{tabular}
\end{table}
%\vspace{-5mm}

As we can observe in Table 2, the concept/irrelevant collocates classification F1-score on the AR dataset is relatively high since the test and training sets are from a similar distribution, in which case the ontology learning system performs very well. In VOQ, since the test data is not from a similar distribution, i.e. the VOQ verbatims are more verbose, the performance on VOQ is much worse than that on AR. The Survey data is conditionally sampled from AR, and therefore is also from a similar distribution as training, which results in good classification performance. Moreover, on AR, the F1-score for each N-gram is 0.88, 0.81, 0.83, 0.86 for $N=1,2,3,4$, respectively. The F1-score for 1-gram is better primarily because we have a polysemy centroid feature to capture polysemy meanings of 1-grams, which very likely have different polysemies. For higher grams, the performance is also very good, and we presume this is because longer concepts are more easily captured by the algorithm while shorter concepts can be easily confused with irrelevant collocates. 

\begin{comment}
\begin{table}[h]\centering
  \begin{tabular}{|l | c |c|c|c|}
  \hline
     \textbf{Dataset} &\textbf{1-gram}&\textbf{2-gram}&\textbf{3-gram}&\textbf{4-gram}  \\ \hline
AR &0.88&0.81& 0.83&0.86\\\hline

  \end{tabular}
      \caption{The N-gram breakdown of concepts and irrelevant collocates classification results on AR.}
\end{table}
\end{comment}

% On average, the GART dataset consists of 37\% less number of stop words and noise words when compared with the VOQ dataset. We do not customize the stop word list and the noise word list specifically to cater to a specific application domain. This results in limited success in terms of reducing the stop words and the noise words from the data. 
\begin{table}[h]\centering
      \caption{The evaluation of concept type classification algorithm. }
  \begin{tabular}{|l | c |c|c|}
  \hline
     \textbf{Dataset}  &\textbf{Precision}&\textbf{Recall}&\textbf{F1-score}  \\ \hline
AR &0.82&0.82& 0.82\\\hline
VOQ & 0.84 &0.65&0.73 \\\hline
Survey & 0.82 &0.80&0.81 \\\hline
  \end{tabular}
\end{table}
%\vspace{-5mm}

We follow the same approach to evaluate the performance of the second stage classifier which takes as the input the concepts classified by the first stage and then assigns concept types. The test set sizes are 800, 1,500, 900 for AR, VOQ and Survey, respectively. Note that the `concepts' passed to the second stage classifier could be incorrectly classified by the first stage classifier, i.e. some inputs could be irrelevant collocates. Each irrelevant collocate input to the second stage classifier is counted as falsely predicted regardless of the type predicted by the classifier. Despite of this, as it can be seen from Table 3, the concept type assignment model shows good Precision rate, however, the Recall rate is lower primarily because of the false negative rate, i.e. the classifier misses on assigning types to long phrases. It is important to note that although the VOQ dataset is generated from a completely different data source, the second stage classifier shows a very good performance. 

\begin{figure}[h]
    \includegraphics[width=0.4\textwidth]{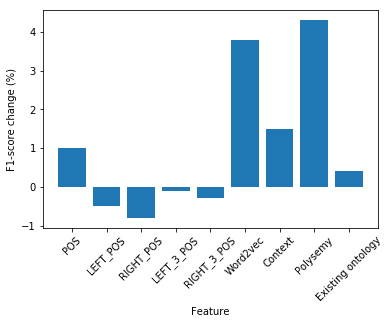}
    \centering
  \caption{Change of F1-score when dropping each feature.}
  \centering
  \end{figure}

Next, we calculate feature importance by recording how much F1-score drops when we remove each feature. The higher the value, the more important the feature. As we can see in Figure 3, the features that contribute most to the F1-score are Word2vec, Context, Polysemy and POS, which is consistent with our observation in backward elimination algorithm. The two most important features are Polysemy (4.3\%) and Word2vec (3.8\%), which shows the significance of applying word embeddings in ontology learning.

\begin{table}[h]
\centering
       \caption{Examples of classification results, where `NONE' denotes irrelevant collocates.}  
\begin{tabular}{|l | l  | l |}
  \hline
     \textbf{COLLOCATE}  & \textbf{PREDICTED} & \textbf{TRUE TYPE}  \\ \hline
RECOVER&A&A\\\hline
NO POWER PUSHED&NONE & NONE\\\hline
HIGH MOUNT BRAKE BULB&B & B\\\hline
PARK LAMP &NONE & B\\\hline
 ROUGH IDLE RIGHT SIDE
 & B& NONE\\\hline
ENGINE CUTS OFF
&NONE & C\\\hline
  \end{tabular}
\end{table}
%\vspace{-5mm}

Table 4 shows typical examples of the correctly and incorrectly classified concepts and irrelevant collocates. Note that only the collocates are shown in Table 4 but the features are not shown. There are some critical reasons that are identified to contribute to the misclassification. First, the POS tags associated with each collocate considered during the training stage is one of the crucial features and it turns out that POS tags assigned by the POS tagger on our data are inconsistent sometimes. For example, in `PARK LAMP,' the POS tagger tags it as `VBN NNP,' while it should be tagged as `NNP NNP' since `PARK' here is not a verb. Second, the real-world data comes in different flavors in terms of the stop and noise words. While standard English stop words and noise words allow us to reduce the non-descriptive collocates in the data, we need a more comprehensive stop and noise word customized dictionary. Moreover, such a dictionary needs to be a living verbatim that requires timely augmentation to ensure as complete coverage to such words as possible. For example, `OFF,' which is usually regarded as a stop word in English, should not be in our customized stop words list since collocates such as `ENGINE CUTS OFF' need `OFF' for grammatical accuracy. Third, collocates that are combinations of types are usually confusing. In our data, collocates such as `ENGINE CUTS OFF' consist of two classes fused together, i.e. collocate `ENGINE' is B, while collocate `CUTS OFF' is C. To handle such cases, we need to have more representatives within the training dataset.

\section{Conclusion}

We propose a two-stage classification system for automatically learning an ontology from unstructured text data. The proposed framework initially cleans the noisy data by correcting different types of noises observed in verbatims. The corrected text is then passed through our two-stage classifier. In the first stage, the classification algorithm automatically classifies collocates into concepts and the irrelevant collocates. Next, the concepts extracted are provided as the input to the second stage classification algorithm which automatically assigns further types to them. In our approach, we not only use the surface features observed in the data, e.g. POS, but we also apply latent features such as word embeddings and polysemy features associated with collocates. As shown in the evaluation, the combination of surface features together with latent features provides necessary discrimination to correctly classify collocates. Our system has been successfully deployed as a proof of concept in a real world domain. %In the future, our aim is to 1. develop a deep neural network to classify the phrases, 2. extend our existing system to extract longer N-grams, such as 5-grams, 6-grams, and 7-grams to ensure additional coverage to extract the ontology, and 3. as described earlier, we also wish to make a complete stop words and noise words lists to further reduce the noises.
%\section{Acknowledgement}

%\bibliographystyle{splncs04}
\end{document}